\documentclass[superscriptaddress,prd,aps,showpacs,nofootinbib,showkeywords,eqsecnum,preprint]{revtex4-2}

\usepackage{graphicx,color,amsmath,amsxtra}
\usepackage{epsf}
\usepackage{amssymb}
\usepackage{enumerate}
\usepackage{hhline}
\usepackage{array}
\usepackage{tabularx}
\usepackage[unicode]{hyperref}
\usepackage{graphicx}                
\usepackage{epstopdf}

\begin{document}
\pagestyle{myheadings}
\title{Effect of $R^2$ on the stability of de Sitter solution of the generalized Einsteinian cubic gravity}
\author{Tuan Q. Do}
\email{tuan.doquoc@phenikaa-uni.edu.vn}
\affiliation{Phenikaa Institute for Advanced Study, Phenikaa University, Hanoi 12116, Vietnam}
\date{\today} 

\begin{abstract}
In this paper, we would like to investigate whether a generalized Einsteinian cubic gravity, in which three possible cubic interactions ${\cal P}$, ${\cal C}$, and ${\cal C}'$ are treated on an equal footing,  admits a de Sitter solution as its stable cosmological solution. As a result, we are able to confirm the existence of the corresponding de Sitter solution for this gravity by solving analytically its field equations. Remarkably, only the cubic interaction ${\cal P}$ gives rise to the existence of the de Sitter solution. Then, we convert the field equations into the corresponding dynamical system for a stability analysis purpose. A fixed point of this dynamical system is found and shown to be equivalent to the obtained de Sitter solution. However, the perturbed dynamical system turns out to be incomplete, leaving undetermined information of the stability of the fixed point (or equivalently the de Sitter solution). Fortunately, we show that this loophole can be cured once the well-known Starobinsky term $R^2$ is introduced into the action of the generalized Einsteinian cubic gravity, despite the fact that it contributes nothing to the value of the de Sitter solution. 
\end{abstract}
\maketitle
\newpage
%%%%%%%%%%%%%%%%%%%%%%%%%%%%%%%%%%%%%%%%%%%%%%%%%%%%%%
\section{Introduction} \label{intro}
Going beyond the pure Einstein's gravity has been one of the most promising approaches to explain the nature of the leading cosmological observations such as the accelerated expansion of late-time universe, which cannot be realized within the pure Einstein gravity \cite{Copeland:2006wr,Sotiriou:2008rp,DeFelice:2010aj,Carroll:2004de,Amendola:2006we,Nojiri:2010wj,Nojiri:2017ncd}. Besides, it has been expected that the origin of the cosmic inflation \cite{Starobinsky:1980te,Guth:1980zm,Linde:1981mu,Linde:1983gd} could be revealed in terms of geometrical deviations from the pure Einstein's gravity involving only the Ricci scalar $R$ \cite{Nojiri:2010wj,Nojiri:2017ncd}. One of the main motivations for this expectation is due to the success of the well-known Starobinsky inflationary model \cite{Starobinsky:1980te}, whose predictions have fitted very well with the latest observational data of the Planck probe \cite{Akrami:2018odb}. The key feature of the Starobinsky model is associated with an introduction of the Ricci scalar squared term, i.e., $R^2$, which is nothing but the leading quantum correction \cite{Starobinsky:1980te}. For a recent discussion on the cosmological aspects of the Starobinsky model, see Ref. \cite{Ketov:2025nkr}. 

However, a very recent data of the Atacama Cosmology Telescope (ACT) seems to slightly disfavor the Starobinsky model \cite{AtacamaCosmologyTelescope:2025nti}. In particular, its observed spectral index $n_s$ seems to be slightly larger than a  theoretical value predicted by the Starobinsky model \cite{AtacamaCosmologyTelescope:2025nti}. Very soon after the data release of ACT, many modifications/extensions of the Starobinsky model have been proposed, e.g., see Refs. \cite{Addazi:2025qra,Ketov:2025cqg,Ellis:2025ieh} for example, in order to maintain its legacy. Historically, the Starobinsky model has been extended in various directions, mostly due to theoretical sides, e.g., see Refs. \cite{Myrzakulov:2014hca,Rodrigues-da-Silva:2021jab,Ivanov:2021chn,Ketov:2022zhp,Do:2023yvg,Pham:2024fub,Kamenshchik:2024kay,Toyama:2024ugg,preprint} for some recent relevant works. A common point can be easily found in these extensions is that they admit no stable de Sitter inflationary solution. This result seems to be in harmony with the fact that the Starobinsky model only admits a quasi-de Sitter solution as its attractor. Additionally,  an exact de Sitter solution is prohibited to exist in the Starobinsky model. 

Remarkably, the existence of de Sitter inflationary solutions along with their stability seem to play an important criteria for  seeing whether a modified gravity model can be suitable for describing the inflationary phase of early universe \cite{Faraoni:2004dn,Faraoni:2005vk,Toporensky:2006kc,Elizalde:2014xva,Pozdeeva:2019agu,Vernov:2021hxo}. For example, the authors of papers shown in Refs.  \cite{Elizalde:2014xva,Pozdeeva:2019agu,Vernov:2021hxo} have argued that admitting unstable exact de Sitter solutions or no exact de Sitter ones could be a smoking gun for a gravity model to be a realistic inflationary model. This is due to the fact that  we will not face to the so-called eternal inflation issue, which probably leads to a multiverse scenario \cite{Guth:2007ng}, if stable (exact) de Sitter inflationary solutions are absent. In case that a stable de Sitter inflationary exists, the so-called graceful exit is supposed to appear in order to avoid the eternal inflation phase, smoothly connecting the early universe to a late-time expanding universe \cite{Brustein:1994kw}.

In cosmology, the Starobinsky model \cite{Starobinsky:1980te} can be classified as a fourth-order gravity since its field equations are fourth-oder ordinary differential equations (ODE) \cite{Whitt:1984pd,Maeda:1987xf,Barrow:1988xh,Schmidt:2006jt}. It has been expected  for a long time that higher-order gravities could be renormalizable and therefore could be relevant to quantum gravity \cite{Stelle:1976gc}.   On the other hand, the Starobinsky model can be called a quadratic gravity since it involves the quadratic term of the Ricci scalar, i.e., $R^2$ \cite{Barrow:2006xb,Alvarez-Gaume:2015rwa,Salvio:2018crh,Asorey:2024oxw}. Besides the inflationary aspect shown above, the Starobinsky model is a very unique fourth-order gravity since it is free from the so-called Ostrogradsky ghost, which could emerge from higher-order terms, such as the third-order and fourth-order ones, in  field equations \cite{Woodard:2015zca}.  It is worth noting that  the Ostrogradsky ghost in ghostful fourth-order gravities should have a mass sufficiently heavier than the energy scale we are being interested in, such as the inflation scale if we would like to consider the inflationary universe, in order to ensure the predictability of the gravities. Otherwise, the considered gravities would be out of control due to an instability caused by the Ostrogradsky ghost \cite{Aoki:2019snr,Lambiase:2025qyl}.

 Due to the ghost-free property, one might think of a possibility that  the Starobinsky model and similar gravity models could act as second-order gravities through a suitable field redefinition. Interestingly, such a thing has been done by many people, e.g., see Refs. \cite{Whitt:1984pd,Barrow:1988xh,Strominger:1984dn}, in which the Starobinsky model is conformally transformed into the Einstein gravity plus an effective scalar field, whose equations of motion are just the second-order ODEs. Another interesting example can be seen in Ref. \cite{Ma:2017jcy}, in which the author has used an effective  technique based on the Legendre transformation to convert a fourth-order gravity involving $R^2$ into a second-order gravity plus additional field(s).

Naturally,  one might ask if there are other higher-order gravities, whose field equations are still of fourth-order ODE. Remarkably,  there have been some interesting works in literature concerning this question. For example, the so-called Einsteinian cubic gravity (ECG) has been proposed in Ref. \cite{Bueno:2016xff} then generalized in some follow-up papers \cite{Hennigar:2017ego,DeFelice:2023vmj}. There is no doubt that the ECG as well as its generalizations have received a lot of attention recently due to their novelty \cite{Hennigar:2016gkm,Bueno:2016ypa,Arciniega:2018fxj,Arciniega:2018tnn,Cano:2020oaa,Bueno:2016lrh,Bueno:2017sui,Hennigar:2018hza,Pookkillath:2020iqq,Bueno:2023jtc}. In harmony with these interesting works, we would like to investigate the existence along with the stability of de Sitter solution within a generalized Einsteinian cubic gravity (GECG), which was firstly considered in Ref. \cite{DeFelice:2023vmj}. As a result, an exact de Sitter solution has been found to the GECG and will be presented in Sect. \ref{sec3}. However, its stability remains unclear due to its perturbed dynamical system is not complete. This incompleteness is due to the existence of some mixed terms of second- and fourth-order time derivatives in field equation.  Fortunately, this loophole can be cured once the well-known Starobinsky term $R^2$ is introduced into the action of the generalized Einsteinian cubic gravity, despite the fact that it contributes nothing to the value of the de Sitter solution. In particular, we will examine the effect of $R^2$ on the stability of the obtained de Sitter solution by considering a Starobinsky-generalized Einsteinian cubic gravity (SGECG). As a result, although $R^2$ does not modify the value of the found de Sitter solution but it plays the leading role in achieving a complete perturbed dynamical system. Interestingly, the obtained de Sitter solution of the SGECG turns out to be unstable as expected for fourth-order gravities. Consequently, one might come to a conclusion, according to Refs. \cite{Elizalde:2014xva,Pozdeeva:2019agu,Vernov:2021hxo}, that the SGECG is more suitable for the early inflationary universe than for the late-time expanding one. 

Interestingly, we have found another approach to resolve the incompleteness issue mentioned above by turning off the higher-order time derivatives of the GECG, so that the corresponding field equations become second ODEs. This thing can be done with a proper setting of field parameters.  In this case, the de Sitter solution, whose value is identical to that found for the full GECG, turns out to be stable as expected for second-order gravities.  Hence, this special case becomes more relevant to the late-time expanding universe. 

The present paper will be organized as follows: (i) Its brief introduction has been written in Sect. \ref{intro}. (ii) Basic setup of the GECG will be presented in Sect. \ref{sec2}. (iii) Exact de Sitter solution of the GECG and its stability will be investigated in Sect. \ref{sec3}. (iv) The effect of $R^2$ on the stability of the obtained de Sitter solution will be analyzed in Sect. \ref{sec4}. (v) Finally, main results of the present paper will be highlighted in Sect. \ref{final}. Some additional calculations will be presented in the Appendices \ref{app} and \ref{app2} just for verifying the validity of the obtained results.
 %%%%%%%%%%%%%%%%%%%%%%%%%%%%%
\section{Generalized Einsteinian cubic gravity} \label{sec2}
\subsection{Action}
Let us begin this section by recalling an action of the GECG \cite{DeFelice:2023vmj},
\begin{equation}
S_{\rm GECG} =\frac{1}{16\pi G}\int d^4 x \sqrt{-g} \left[R+\alpha {\cal P} +\kappa {\cal C} +\mu {\cal C}' \right],
\end{equation}
where $\alpha$, $\kappa$, and $\mu$ are all constant. In addition, three cubic interactions are defined as follows
\begin{align}
{\cal P} &=12 R_a{}^c{}_b{}^d R_c{}^e{}_d{}^f R_e{}^a{}_f{}^b +R_{ab}{}^{cd} R_{cd}{}^{ef} R_{ef}{}^{ab} -12 R_{abcd}R^{ac}R^{bd}+8R_a{}^b R_b{}^cR_c{}^a,\\
{\cal C}& = R_{abcd}R^{abc}{}_e R^{de} -\frac{1}{4} R_{abcd}R^{abcd}R-2R_{abcd}R^{ac}R^{bd} +\frac{1}{2}R_{ab}R^{ab}R,\\
{\cal C}' &= R_a{}^bR_b{}^cR_c{}^a-\frac{3}{4}R_{ab}R^{ab}R +\frac{1}{8}R^3.
\end{align}
It appears that the first cubic interaction, i.e., ${\cal P}$, was derived in the so-called Einsteinian cubic gravity (ECG) \cite{Bueno:2016xff}, while the second cubic interaction, i.e., ${\cal C}$, was considered later in the so-called cosmological Einsteinian cubic gravity (CECG) \cite{Arciniega:2018fxj} for investigating cosmic inflation. The last cubic interaction, i.e., ${\cal C}'$, has been introduced in a recent paper \cite{DeFelice:2023vmj} for a completeness. In other words, if we turn off both ${\cal C}$ and ${\cal C}'$, we will have the ECG \cite{Bueno:2016xff} with the following action given by
\begin{equation}
S_{\rm ECG} =\frac{1}{16\pi G}\int d^4 x \sqrt{-g} \left[R+\alpha {\cal P}  \right].
\end{equation}
 On the other hand, if we turn off only ${\cal C}'$, we will have the CECG \cite{Arciniega:2018fxj}, whose action reads 
 \begin{equation}
S_{\rm CECG} =\frac{1}{16\pi G}\int d^4 x \sqrt{-g} \left[R+\alpha {\cal P} +\kappa {\cal C}  \right].
\end{equation}
Different purposes, e.g., studying black holes or cosmic inflation, would correspond to one of three mentioned cubic gravities, i.e., the ECG, CECG, and GECG. See, for example, Refs. \cite{Bueno:2016lrh,Bueno:2016ypa,Bueno:2017sui,Hennigar:2018hza,Pookkillath:2020iqq,Bueno:2023jtc} for recent implications of the ECG as well as its extensions.

It is interesting to mention that if $\kappa= -8\alpha$ is taken then the field equations of CECG will be second-order ones in the Friedmann-Lemaitre-Robertson-Walker (FLRW) background as pointed out in Ref. \cite{Arciniega:2018fxj}. We will come back to this special feature once the general field equations are derived. 
%%%%%%%%%%%%%%%
\subsection{Field equations for the FLRW metric}
Now, we would like to derive the corresponding field equations of the GECG for the spatially flat  FLRW metric,
\begin{equation} \label{metric}
ds^2 =-N^2(t)dt^2 +e^{2\beta(t)} \left(dx^2 + dy^2 +dz^2 \right),
\end{equation}
which is apparently a homogeneous and isotropic spacetime obeying the cosmological principle.  Here, $N(t)$ and $\beta(t)$ are the lapse function and scale factor, respectively.  One might notice that $N(t)$ is always set to be one in usual scenarios. A reason for introducing $N(t)$ in the above metric is that we would like to derive the Friedmann field equation from its Euler-Lagrange (EL) equation rather than from the Einstein field equations, which are basically tensorial equations. This effective method has been used in many published papers, e.g., Refs. \cite{Myrzakulov:2014hca,Do:2023yvg,Pham:2024fub,Asorey:2024oxw,preprint,Do:2020vdc}.  It should be noted that one can add an arbitrary constant $\beta_0$ to $\beta(t)$, i.e., $\beta(t) \to \beta(t) +\beta_0$, in the FLRW metric \eqref{metric}. However, this adding seems to be unnecessary since we can always omit $\beta_0$ via a spatial rescaling as $(x,y,z) \to e^{-\beta_0} (x,y,z)$. More importantly, the existence of $\beta_0$ is clearly irrelevant to the so-called Hubble parameter $H(t)=\dot\beta(t)$, which plays an important role in comparisons with observational data.

Before going to derive the corresponding field equations of the GECG, it is necessary to work out its Lagrangian given by
\begin{equation}
{\cal L} = \sqrt{-g} \left[ R +\alpha {\cal P} +\kappa {\cal C} +\mu {\cal C}' \right].
\end{equation}
It appears for the FLRW metric \eqref{metric} that \cite{xAct}
\begin{align}
 \sqrt{-g} & =N e^{3\beta}, \\
  R & = -6N^{-3} \left[ \dot N \dot\beta - N \left(\ddot\beta+2\dot\beta^2 \right) \right],\\
  {\cal P}&= -48N^{-9} \left[\dot N \dot\beta - N\left(\ddot\beta-2\dot\beta^2 \right) \right] \left[\dot N \dot\beta -N \left(\ddot\beta+\dot\beta^2 \right) \right]^2,\\
  {\cal C}&= - 6N^{-9} \left(\dot N \dot\beta - N \ddot\beta \right)^3,\\
  {\cal C}'&= \frac{1}{2}{\cal C},
\end{align}
where $\dot N \equiv dN/dt$, $\dot\beta \equiv d\beta/dt$, and $\ddot\beta \equiv d^2 \beta/dt^2$.
Very interestingly, the result, ${\cal C}' =\frac{1}{2}{\cal C}$, can be seen in Ref. \cite{DeFelice:2023vmj}, in which a static and spherically symmetric background has been considered. Hence, one might expect that the proportion between ${\cal C}'$ and ${\cal C}$ could be a general feature of the GECG for arbitrary backgrounds. 

Provided with the explicit Lagrangian ${\cal L}$ defined above, we are now going to consider the EL equation for $N$ as well as $\beta$. It turns out that the  EL equation for $N$ takes the following form,
\begin{equation}
\frac{\partial {\cal L}}{\partial N} -\frac{d}{dt} \left(\frac{\partial {\cal L}}{\partial \dot N}\right)=0,
\end{equation}
while the EL equation for $\beta$ reads
\begin{equation}
\frac{\partial {\cal L}}{\partial \beta} -\frac{d}{dt} \left(\frac{\partial {\cal L}}{\partial \dot \beta}\right) + \frac{d^2}{dt^2} \left(\frac{\partial {\cal L}}{\partial \ddot\beta}\right)=0.
\end{equation}
It appears that the difference between these two equations lies on the second time derivative in the latter equation. This terms exists because the Lagrangian ${\cal L}$ contains the second time derivative of $\beta$. 

As a result, these two EL equations become, after setting $N=1$,
\begin{align}
\label{field-equation-1}
&2\dot\beta^2 + \alpha \left(16\dot\beta^6 +144 \dot\beta^2 \ddot\beta^2 -32\ddot\beta^3 +96 \dot\beta \ddot\beta \beta^{(3)} \right) + \left( 2\kappa + \mu\right) \left(9\dot\beta^2 \ddot\beta^2 -2 \ddot\beta^3 +6\dot\beta \ddot\beta \beta^{(3)} \right)=0,\\
\label{field-equation-2}
&6\dot\beta^2 +4\ddot\beta +\alpha \left[ 48 \dot\beta^6 +96 \dot\beta^4 \ddot\beta +432 \dot\beta^2 \ddot\beta^2 +192 \ddot\beta^3 +576 \dot\beta \ddot\beta \beta^{(3)} + 96 \left(\beta^{(3)} \right) ^2 +96 \ddot\beta \beta^{(4)} \right] \nonumber\\
& +\left( 2\kappa + \mu\right)  \left[ 27 \dot\beta^2 \ddot\beta^2 +12 \ddot\beta^3 + 36 \dot\beta \ddot\beta \beta^{(3)} + 6\left(\beta^{(3)} \right) ^2  +6 \ddot\beta \beta^{(4)} \right]=0,
\end{align}
respectively. Here, it should be understood as  $\beta^{(3)} \equiv d^3 \beta/dt^3$ and $\beta^{(4)} \equiv d^4 \beta/dt^4$. 

It is apparent that these field equations are nothing but higher-than-second-order ODEs. In particular, the first field equation is the third-order ODE of $\beta$, while the last one is the fourth-order ODE of $\beta$. Due to this result, the ECG, CECG, and GECG can be classified as the fourth-order gravities \cite{Schmidt:2006jt}.  It is straightforward to verify that Eq. \eqref{field-equation-2} can be obtained by combining Eq. \eqref{field-equation-1} with its time derivative. As a result, this relation is just a consequence of the Bianchi identity given by $\nabla^\mu G_{\mu\nu}=0$, with $G_{\mu\nu} \equiv R_{\mu\nu}-\frac{1}{2}Rg_{\mu\nu}$ being the Einstein tensor. In this case, Eq. \eqref{field-equation-1} can be interpreted as the $00$-component of Einstein field equation, while Eq. \eqref{field-equation-2} can be identified with the $ii$-component of  Einstein field equation.  We now have a more evidence to confirm that these field equations, which are derived from the EL equations, are all correct.

 It must be emphasized that although Eq. \eqref{field-equation-2} can be regarded as a differential consequence of the Friedmann equation, i.e., Eq. \eqref{field-equation-1}, we should not ignore it when studying the dynamics of spacetime in order to ensure the validity of all obtained results. Originally, Eq. \eqref{field-equation-2} comes from the $ii$-component of Einstein field equation rather than from the Bianchi identity. Remarkably,  the $ii$-component of Einstein field equation has been used in many gravity models to construct the corresponding dynamical systems, while the $00$-component of Einstein field equation has remained as a constraint equation of the dynamical system, e.g., see Refs.  \cite{Do:2023yvg,Pham:2024fub,Pozdeeva:2019agu,Vernov:2021hxo,Barrow:2006xb}. However, it seems that this approach is not unique. Indeed, it turns out that some other people have only used the $00$-component of Einstein field equation for investigating the dynamics as well as the stability of higher-order gravity models, e.g., see Ref. \cite{Toporensky:2006kc} for an interesting example associated with the quadratic gravity. In this approach, a fourth-order gravity should be renamed as a third-order gravity since its field equation used to study is just a third-order ODE. Additionally, the $ii$-component of Einstein field equation is nothing but a constraint equation.  In harmony with this logic, one may call the pure Einstein's gravity (a.k.a. General Relativity) a first-oder gravity since the $00$-component of its corresponding Einstein field equation is just a first-order ODE. Interestingly, by introducing the Hubble parameter, $H\equiv \frac{\dot a}{a}$, one can even reduce a fourth-order (third-order) gravity of metric scale to a third-order (second-order) gravity of Hubble parameter \cite{Toporensky:2006kc}. More interestingly, the Friedmann equation of Einstein's gravity can be shown to be an algebraic equation of $H$ as stated in Ref. \cite{Toporensky:2006kc}.  Apparently, it seems that the order of gravity models depends on  the preferred variables as well as field equations. To be complete, we will consider the latter approach,  in which only the $00$-component of Einstein field equation is considered, in the Appendix \ref{app}. As expected,  all main results such as de Sitter solutions as well as their stability property obtained in the main sections  will be recovered in this Appendix.

 One might claim that it is quite soon to classify the ECG, CECG, and GECG as the fourth-order gravities due to their field equations derived in the FLRW background metric shown above. To have a general claim, one should consider the tensorial Einstein field equations of these models, in which higher general derivatives of cubic interactions show up and generate fourth-order (spatial and/or time) derivatives of metric components. For example, one might want to see Ref. \cite{Bueno:2016xff} for such a field equation. Since we have limited ourself to the cosmological aspect of the ECG, CECG, and GECG, we have just considered the standard FLRW metric for our analysis. Hence, the ECG, CECG, and GECG have been classified as the fourth-order gravities in the present paper in the sense that they are gravity models, whose field equations involve fourth-order time derivatives. 

Now, we come back to the special case with $\kappa =- 8\alpha$ as mentioned above. It is straightforward to verify the result obtained in Ref. \cite{Arciniega:2018fxj} that the field equations will be second-order ODEs if $\kappa=-8\alpha$, provided that $\mu=0$. In the presence of the non-vanishing $\mu$ (or ${\cal C}'$), we must impose the constraint, $2\kappa+\mu = -16 \alpha$, in order to maintain the second-order field equations for the GECG. However, since we would like to have a general picture, we will keep three parameters $\alpha$, $\kappa$, and $\mu$ independent of each other. Of course, the special case with $2\kappa+\mu = -16 \alpha$ will be discussed later once the general case with $2\kappa+\mu \neq -16 \alpha$ is done. 
%%%%%%%%%%%%
\section{de Sitter solution and its stability}\label{sec3}
\subsection{de Sitter solution}
So far, basic ingredients of the GECG such as its action and field equations for the FLRW background have been presented. We are now going to seek analytical cosmological solutions to the GECG. Among the well-known cosmological solutions, the de Sitter solution has been regarded as the leading one due to its simplicity as well as its compatibility with the cosmological principle, which states that our universe is simply homogeneous and isotropic on large scales ($\gtrsim 100$ Mpc). In fact, de Sitter solutions have been found in many fourth-order gravities, e.g., see Refs. \cite{Ketov:2022zhp,Do:2023yvg,Pham:2024fub,preprint,Toporensky:2006kc,Pozdeeva:2019agu,Vernov:2021hxo,Do:2020vdc} for recent related investigations. 

In order to figure out  the GECG's de Sitter solution, we will consider the following ansatz, which has been used in our previous papers \cite{Do:2023yvg,Pham:2024fub,preprint,Do:2020vdc},
\begin{equation} \label{ansatz}
\beta(t) = \zeta t,
\end{equation}
where $\zeta $ is a constant, which should be positive for an expanding universe. In addition, $\zeta \gg 1$ is necessary for an inflationary universe. It turns out for this choice that
\begin{equation}
\dot\beta = \zeta,\quad \ddot\beta = \beta^{(3)}=\beta^{(4)} =0.
\end{equation}
Consequently, two field equations \eqref{field-equation-1} and \eqref{field-equation-2} will be significantly reduced to the same simple algebraic equation of $\zeta$,
\begin{equation} \label{equation-of-zeta}
8\alpha \zeta^4 +1 =0.
\end{equation}
As a result, a non-trivial solution of  this equation turns out to be
\begin{equation} \label{zeta}
\zeta = \left(-\frac{1}{8\alpha} \right)^{\frac{1}{4}}.
\end{equation}
It now becomes clear that $\alpha$ must be negative definite to ensure the corresponding real value(s) of $\zeta$. On the other hand, only the original term of ECG, i.e., ${\cal P}$, gives rise to the existence of de Sitter solution. The other terms such as  ${\cal C}$ and ${\cal C}'$ contribute nothing to the value of this solution. Apparently, the inflationary constraint $\zeta\gg 1$ leads to $|8\alpha| \ll 1$ or equivalently  $|\alpha| \ll 1/8$. 
%%%%%%%%%%%%%%
\subsection{Dynamical system}
One of the most important properties of all cosmological solutions is their stability against perturbations. Therefore, we would like to investigate in this subsection whether the obtained de Sitter solution is stable or not. Similar to the previous papers \cite{Do:2023yvg,Pham:2024fub,preprint,Do:2020vdc}, our stability analysis follows the powerful method based on the dynamical system, which has been used widely in modern cosmology \cite{Barrow:2006xb,Bahamonde:2017ize}. In particular, we will convert the field equations of the GECG into its corresponding dynamical system in terms of the following dynamical variables defined as \cite{Barrow:2006xb}
 \begin{equation}
B=\frac{1}{\dot\beta^2},\quad Q=\frac{\ddot\beta}{\dot\beta^2},\quad Q_2 =\frac{\beta^{(3)}}{\dot\beta^3}.
 \end{equation}
 As a result, the corresponding autonomous equations, which form the dynamical system of the GECG, are given by
 \begin{align}
  \label{Dyn-1}
 B' &= -2QB,\\
  \label{Dyn-2}
 Q' &=Q_2 -2Q^2,\\
 \label{Dyn-3}
 Q_2 ' & =  \frac{\beta^{(4)}}{\dot\beta^4}-3Q Q_2.
 \end{align}
Here, the prime stands for a derivative w.r.t. a dynamical time variable $\tau$ defined as $\tau =\int \dot\beta dt$. For example, $B' \equiv dB/d\tau$ and so on. Apparently,  the term $ \frac{\beta^{(4)}}{\dot\beta^4}$ appearing in the Eq. \eqref{Dyn-3} can be determined from the field equation \eqref{field-equation-2}. Indeed, one can easily rewrite Eq. \eqref{field-equation-2} in terms of the introduced dynamical variables as follows
 \begin{align} \label{Dyn-4}
& 6B^2 +4QB^2 +\alpha \left(48+96Q+432Q^2 +192 Q^3 +576QQ_2 +96Q_2^2+96 Q\frac{\beta^{(4)}}{\dot\beta^4} \right) \nonumber\\
 &+ \left(2\kappa+\mu \right) \left( 27Q^2+12Q^3 +36QQ_2 +6Q_2^2 +6Q \frac{\beta^{(4)}}{\dot\beta^4}\right)=0.
 \end{align}
 If $Q\neq 0$ as well as $2\kappa+\mu \neq -16\alpha$, we can obtain, from this equation, an explicit expression of $ \frac{\beta^{(4)}}{\dot\beta^4}$ in terms of the dynamical variables given by
 \begin{align} \label{Dyn-7}
 \frac{\beta^{(4)}}{\dot\beta^4} =&~ \frac{-1}{6Q\left(16\alpha+2\kappa+\mu \right)} \left[6B^2 +4QB^2 +\alpha \left(48+96Q+432Q^2 +192 Q^3 +576QQ_2 +96Q_2^2 \right) \right. \nonumber\\
 & \left. + \left(2\kappa+\mu \right) \left( 27Q^2+12Q^3 +36QQ_2 +6Q_2^2 \right) \right].
 \end{align}
 However, we must be very careful if either $Q=0$ or $2\kappa+\mu= -16\alpha$. In other words, the above expression of $\frac{\beta^{(4)}}{\dot\beta^4} $ will no longer be valid for either the vanishing $Q$ or the special case with $2\kappa+\mu= -16\alpha$, in which the field equations are only second-order ODEs.
 We should not forget an important constraint equation, which is nothing but the Friedmann equation  \eqref{field-equation-1} written in terms of the dynamical variables,
 \begin{equation}
 \label{Dyn-5}
 2B^2+\alpha \left(16 +144 Q^2 -32 Q^3 +96 Q Q_2  \right) +\left(2\kappa+\mu \right) \left(9Q^2 -2Q^3 +6 Q Q_2 \right)=0.
 \end{equation}
Indeed, all found fixed points of the dynamical system must fulfill this equation. 
 %%%%%%%%%%%
 \subsection{Fixed points} \label{fixedpoint}
 So far, we have successfully derived the dynamical system of three autonomous equations \eqref{Dyn-1}, \eqref{Dyn-2}, and \eqref{Dyn-3}, with the constraint equations \eqref{Dyn-4} and \eqref{Dyn-5} coming from the field equations of the GECG. In principle, Eqs. \eqref{Dyn-1}, \eqref{Dyn-2}, and \eqref{Dyn-3} remain the same for different fourth-order gravities in the FLRW metric, while the associated constraint equations like Eqs. \eqref{Dyn-4} and \eqref{Dyn-5} will depend on the specific structure of these gravities. In other words, the difference between dynamical systems of fourth-order gravities will be determined by the corresponding field equations.  
 
 Next step of our stability analysis is to seek fixed points to the dynamical system. Mathematically, these fixed points are solutions of the following set of equations,
 \begin{equation}
 B'=Q'=Q_2' =0.
 \end{equation}
 As a result, the first equation, $B'=0$, implies two possibilities, according to Eq. \eqref{Dyn-1},
 \begin{equation}
 Q=0 \quad {\text {or}} \quad B=0.
 \end{equation}
 However, in order to be compatible with the de Sitter solution found in the previous subsection, $B$ should not vanish. Hence, $Q=0$ (or equivalently $\ddot\beta=0$) turns out to be the desired solution. As a result, the second equation, $Q'=0$, implies
 \begin{equation}
 \label{relation}
 Q_2=2Q^2 =0.
 \end{equation}
 The vanishing of $Q_2$ is equivalent to $\beta^{(3)}=0$.
 Finally, the last equation, $Q_2'=0$, gives
 \begin{equation}
  \frac{\beta^{(4)}}{\dot\beta^4}=0.
 \end{equation} 
 It is straightforward to check that the solution $\frac{\beta^{(4)}}{\dot\beta^4}=Q_2=Q=0$ lead both Eqs. \eqref{Dyn-4} and \eqref{Dyn-5} to the same equation of $B$ given by
 \begin{equation}
 B^2+8\alpha=0,
 \end{equation}
 which can be solved to give
 \begin{equation}
 \dot\beta = \left(-\frac{1}{8\alpha}\right)^{\frac{1}{4}}.
 \end{equation}
 Integrating out this equation leads to the de Sitter solution found above,
 \begin{equation}
 \beta =\zeta t,
 \end{equation}
 where $\zeta$ has been defined in Eq. \eqref{zeta}. This result clearly indicates that our obtained fixed point with $B = \sqrt{-8\alpha} \neq 0$ and $\frac{\beta^{(4)}}{\dot\beta^4}=Q_2=Q=0$ is indeed equivalent to the de Sitter solution found above.  It must emphasize that the solution $\frac{\beta^{(4)}}{\dot\beta^4}=Q_2 =0$ has been derived just from the equations $B'=Q'=Q_2' =0$ without the help of Eq. \eqref{Dyn-7}. However, the simultaneously vanishing of both quantities $\frac{\beta^{(4)}}{\dot\beta^4}$  and $Q_2$ should not happen, according to  Eq. \eqref{Dyn-7}. Furthermore, as will be shown below it will lead us to a serious issue of the stability investigation of the de Sitter fixed point.
 
 %%%%%%%%%%%%%%%
 \subsection{Stability of the de Sitter fixed point} 
 Now, we would like to examine whether the obtained fixed point is stable or not against perturbations. Due to the equivalence, both the found fixed point and de Sitter solution should share the same stability property. For convenience, we will call the found fixed point the de Sitter fixed point. As a standard procedure, we first perturb the dynamical variables around this fixed point,
 \begin{equation}
 B \to B+\delta B, \quad Q\to Q+\delta Q, \quad Q_2 \to Q_2+\delta Q_2.
 \end{equation}
 Consequently, the autonomous equations will be perturbed as
 \begin{align}
 \delta B' &= -2 B\delta Q,\\
 \delta Q'&= \delta Q_2 ,\\
 \label{pert-3}
 \delta Q_2'&= \delta \left(\frac{\beta^{(4)}}{\dot\beta^4}\right).
 \end{align}
 The next step should be done is to determine $\delta \left(\frac{\beta^{(4)}}{\dot\beta^4}\right)$ in terms of the other perturbations of dynamical variables  using Eq. \eqref{Dyn-4}.  We observe that $\frac{\beta^{(4)}}{\dot\beta^4}$  always couples with $Q$ in Eq. \eqref{Dyn-4}, forming their two mixed terms. Unfortunately, these terms will cause a serious issue in determining explicitly  $\delta \left(\frac{\beta^{(4)}}{\dot\beta^4}\right)$. In particular, it turns out for the de Sitter fixed point with $\frac{\beta^{(4)}}{\dot\beta^4}=Q=0$ that 
 \begin{equation}
 \delta \left(Q \frac{\beta^{(4)}}{\dot\beta^4} \right) = Q \delta \left( \frac{\beta^{(4)}}{\dot\beta^4} \right)+ \left( \frac{\beta^{(4)}}{\dot\beta^4} \right)\delta Q =0 .
 \end{equation}
 This result implies that we will no longer have any term having $\delta \left( \frac{\beta^{(4)}}{\dot\beta^4} \right)$ in the perturbed equations. In other words, we cannot derive $\delta \left(\frac{\beta^{(4)}}{\dot\beta^4}\right)$ from  Eq. \eqref{Dyn-4}  in order to complete Eq.  \eqref{pert-3}. This result means that Eq. \eqref{pert-3} is indeed incomplete and $\delta Q_2'$ will therefore  be unconstrained. Hence, the stability of the de Sitter fixed point cannot be analyzed due to the lack of information of $\delta \left(\frac{\beta^{(4)}}{\dot\beta^4}\right)$ as well as $\delta Q_2'$. Again, we must emphasize that the main reason is due to the mixed terms between $Q$ and $\frac{\beta^{(4)}}{\dot\beta^4} $ in Eq.  \eqref{Dyn-4}.  To ensure the validity of our conclusion, we will perform an alternative stability analysis, which is based on a different method used in some published papers, e.g., Ref. \cite{CamposDelgado:2024jst}, as a cross-check. Detailed calculations and discussions will be presented in the Appendix \ref{app2}.
  
 Fortunately, this issue could be resolved if we introduce an additional term involving only $\frac{\beta^{(4)}}{\dot\beta^4}$ into Eq. \eqref{Dyn-4}. Our past experiences in studying fourth-order gravities \cite{Do:2023yvg,Pham:2024fub,preprint} suggest us that the Starobinsky term $R^2$ is exactly a desired resolution. Therefore, we will consider a Starobinsky-GECG  scenario in the next section for stability reasons. 
 %%%%%%%%%%%%%%%
 \subsection{Special case}
 In this subsection, we will focus on the special case mentioned above, in which the field equations of GECG will only be second-order ODEs thanks to the special setting $2\kappa+\mu =-16\alpha$. As a result, both general field equations \eqref{field-equation-1} and \eqref{field-equation-2} will be reduced to 
 \begin{align}
\label{field-equation-special-1}
&2\dot\beta^2 +16 \alpha  \dot\beta^6=0,\\
\label{field-equation-special-2}
&6\dot\beta^2 +4\ddot\beta +\alpha \left( 48 \dot\beta^6 +96 \dot\beta^4 \ddot\beta \right)=0,
\end{align}
where all terms containing $\beta^{(3)}$ and $\beta^{(4)}$ disappear accordingly. It turns out that Eq. \eqref{field-equation-special-2} can be derived directly from Eq.  \eqref{field-equation-special-1} by taking the time derivative of both its sides. Furthermore, these equations are identical to Eqs. (6) and (7) derived in Ref. \cite{Arciniega:2018fxj}, respectively, in the vacuum limit with a vanishing cosmological constant, provided that $H=\dot\beta$ and $\dot H =\ddot\beta$.  It is straightforward to see that these field equations always admit the de Sitter solution found above as their non-trivial cosmological solution. 

For the stability analysis, we only need  the dynamical variable $B$ since the field equations are all second-order ODEs. As a result, the corresponding dynamical system is formed by its autonomous equation given by
\begin{equation}
B'=2 B \frac{6B^2+48\alpha}{4B^2+96\alpha},
\end{equation}
with the help of the field equation \eqref{field-equation-special-2}.
 Consequently, the corresponding non-trivial fixed point with $B\neq 0$ is found to be
 \begin{equation}
 B^2 =-8\alpha,
 \end{equation}
 which is exactly identical to the de Sitter fixed point shown above, confirming our earlier claim. This solution satisfies the field equation \eqref{field-equation-special-1}. 
 \begin{figure}[hbtp] 
 \centering
	  \includegraphics[scale=0.7]{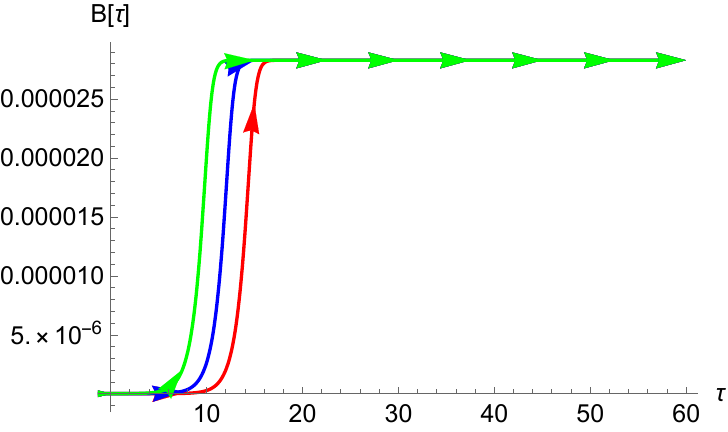}
	
\caption{\it The de Sitter fixed point of the special case of the GECG acts as an attractor. The parameters have been chosen as $\alpha=-10^{-10}$ and $M_p=1$ such that the value of the de Sitter fixed point is $B \simeq 2.82843 \times 10^{-5}$. Different colors correspond to different initial conditions. In particular, the red, blue, and green curves correspond to $B[0]=10^{-11}$, $B[0]= 10^{-10}$, and $B[0]=10^{-9}$, respectively. }
\label{fig1}
\end{figure}

  Perturbing the autonomous equation around this de Sitter fixed point yields
 \begin{equation}
 \delta B' = -3\delta B.
 \end{equation}
 Solving this equation implies a non-trivial solution of perturbation,
 \begin{equation}
 \delta B \propto e^{-3\tau},
 \end{equation}
 which will decay exponentially as $\tau$ becomes large, ensuring the stability of the corresponding de Sitter solution. In addition, numerical solutions shown in Fig. \ref{fig1} clearly indicate that the de Sitter fixed point is an attractor. One can therefore judge, following discussions in Refs. \cite{Elizalde:2014xva,Pozdeeva:2019agu,Vernov:2021hxo}, that this special case of the GECG, whose field equations are the second-order ODEs, turns out to be more suitable for the late-time expanding universe than for the early inflationary universe.
   %%%%%%%%%%%%%%%%%
 \section{Starobinsky-GECG} \label{sec4}
 As stated above, we would like to propose in this section an action of Starobinsky-GECG (SGECG) model given by
 \begin{equation} \label{SGECG}
 S_{\rm SGECG} =\frac{1}{16\pi G}\int d^4 x \sqrt{-g} \left[R+\gamma R^2 +\alpha {\cal P} +\kappa {\cal C} +\mu {\cal C}' \right],
\end{equation}
where $\gamma$ is an additional parameter, which must be positive definite as required in the pure Starobinsky gravity \cite{Starobinsky:1980te,Ketov:2025nkr}. As a result, the corresponding field equations of the SGECG defined in terms of the EL equations are given by
\begin{align}
\label{field-equation-3}
&2\dot\beta^2 +\gamma \left(72\dot\beta^2 \ddot\beta -12\ddot\beta^2 +24\dot\beta \beta^{(3)} \right) + \alpha \left(16\dot\beta^6 +144 \dot\beta^2 \ddot\beta^2 -32\ddot\beta^3 +96 \dot\beta \ddot\beta \beta^{(3)} \right) \nonumber\\
& +\left( 2\kappa + \mu\right) \left(9\dot\beta^2 \ddot\beta^2 -2 \ddot\beta^3 +6\dot\beta \ddot\beta \beta^{(3)} \right)=0,\\
\label{field-equation-4}
&6\dot\beta^2 +4\ddot\beta +\gamma \left(216\dot\beta^2 \ddot\beta +108 \ddot\beta^2+144 \dot\beta \beta^{(3)} +24 \beta^{(4)} \right) \nonumber\\
&+\alpha \left[ 48 \dot\beta^6 +96 \dot\beta^4 \ddot\beta +432 \dot\beta^2 \ddot\beta^2 +192 \ddot\beta^3 +576 \dot\beta \ddot\beta \beta^{(3)} + 96 \left(\beta^{(3)} \right) ^2 +96 \ddot\beta \beta^{(4)} \right] \nonumber\\
& +\left( 2\kappa + \mu\right)  \left[ 27 \dot\beta^2 \ddot\beta^2 +12 \ddot\beta^3 + 36 \dot\beta \ddot\beta \beta^{(3)} + 6\left(\beta^{(3)} \right) ^2  +6 \ddot\beta \beta^{(4)} \right]=0.
\end{align}
As expected, it is straightforward to check that Eq. \eqref{field-equation-4} can be obtained from combining Eq. \eqref{field-equation-3} and its time derivative. In other words, one can state that Eq. \eqref{field-equation-4} is nothing but a differential consequence of Eq. \eqref{field-equation-3}. And again, this result does confirm the validity of these two field equations. 
Very interestingly, all terms coming from the quadratic term $R^2$ will disappear for the ansatz \eqref{ansatz}. It means that we still obtain the same de Sitter solution derived above for the SGECG scenario since the Starobinsky term $R^2$ does not contribute any additional term into Eq. \eqref{equation-of-zeta}. Similar situations can be seen in our previous studies on some extensions of Starobinsky gravity \cite{Do:2023yvg,Pham:2024fub,preprint}

Now, we would like to revisit the stability issue. As a result, it is safety to have the following relation,
\begin{align}
 \frac{\beta^{(4)}}{\dot\beta^4} =&~ \frac{-1}{6\left[ 4\gamma B+ \left(16\alpha +2\kappa +\mu\right) Q  \right]} \left[6B^2 +4QB^2 +\gamma B\left(216 Q+108Q^2+144Q_2 \right) \right. \nonumber\\
 &\left. +\alpha \left(48+96Q+432Q^2 +192 Q^3 +576QQ_2 +96Q_2^2 \right) \right. \nonumber\\
 & \left. + \left(2\kappa+\mu \right) \left( 27Q^2+12Q^3 +36QQ_2 +6Q_2^2 \right) \right],
 \end{align}
 even when $Q=0$ or $2\kappa+\mu =-16\alpha$. Since the de Sitter fixed point with $B = \sqrt{-8\alpha} \neq 0$ and $\frac{\beta^{(4)}}{\dot\beta^4}=Q_2=Q=0$ found above is still valid for the SGECG, we have the following perturbed dynamical system of the SGECG,
 \begin{align}
 \delta B' &= -2 B\delta Q,\\
 \delta Q'&= \delta Q_2 ,\\
 \label{pert-4}
 \delta Q_2'&= \frac{-1}{24\gamma B} \left[12B\delta B +4B^2 \delta Q +216 \gamma B \delta Q +144\gamma B \delta Q_2 +96\alpha \delta Q  \right].
 \end{align}
 Furthermore, if we take an additional perturbed equation from the constraint Eq. \eqref{field-equation-3},
 \begin{equation}
 4B\delta B +72\gamma B \delta Q +24\gamma B\delta Q_2 =0,
 \end{equation}
 then Eq. \eqref{pert-4} will be simplified as
 \begin{equation}
 \delta Q_2' = - \frac{8\alpha}{3\gamma B}\delta Q -3\delta Q_2.
 \end{equation}
 Taking exponential perturbations, 
 \begin{equation} \label{expo-pertub}
 \delta B = A_B e^{\lambda \tau}, \quad \delta Q = A_Q e^{\lambda \tau}, \quad \delta Q_2 = A_{Q_2} e^{\lambda \tau}, 
 \end{equation}
 the above perturbed equations can be written as a set of homogenous algebraic equations, whose matrix form is given by
 \begin{equation} \label{stability-equation}
 {\cal M}\left( {\begin{array}{*{20}c}
   A_B  \\
   A_Q  \\
   A_{Q_2} \\
 \end{array} } \right) \equiv \left[ {\begin{array}{*{20}c}
   {\lambda} & {2B} & {0 }   \\
   { 0} & {\lambda} & {-1 }  \\
     {0  } & { \frac{8\alpha}{3\gamma B}} & {  \lambda+3 }  \\
 \end{array} } \right]  \left( {\begin{array}{*{20}c}
   A_B  \\
   A_Q  \\
   A_{Q_2} \\
 \end{array} } \right) = 0.
\end{equation}
It is well known that this set of homogeneous equations admits non-trivial solutions if and only if 
\begin{equation}
\det {\cal M}=0,
\end{equation}
which can be explicitly expanded to be an equation of $\lambda$ such as
\begin{equation} \label{root}
\lambda \left(3\gamma B \lambda^2 +9\gamma B \lambda +8\alpha \right)=0.
\end{equation}
Besides a trivial root $\lambda_1=0$, this equation admits two non-trivial ones,
\begin{equation} \label{lambda-pm}
\lambda_{2,3} = \lambda_{\pm}= - \frac{1}{6}\left(9 \pm \sqrt{81 -\frac{96\alpha}{\gamma B}} \right).
\end{equation}
Since $\gamma >0$ as required in the pure Starobinsky gravity and $\alpha<0$ due to the existence of the de Sitter solution, it becomes clear that $-\frac{96\alpha}{\gamma B} >0$. Therefore, it turns out that $ \lambda_3 =\lambda_- >0$.  Mathematically, the positivity of $\lambda_3$ will definitely make the found fixed point unstable since $e^{\lambda \tau}$ will blow up as $\tau \gg 1$. If we change the sign of $\gamma$ from positive definite to negative definite, then we will no longer have any positive root $\lambda$ of Eq. \eqref{root}, meaning that the corresponding de Sitter solution, whose value remains the same, will be stable. It now becomes clear that the $R^2$ term does not contribute to the value of the de Sitter solution but it does affect on the stability of this solution. Remarkably, the similar situation can be found in the so-called Starobinsky-Bel-Robinson gravity \cite{Do:2023yvg}, in which the Bel-Robinson correction gives rise the existence of de Sitter solution but the Starobinsky term determines the stability of this solution.  

\begin{figure}[hbtp] 
 \centering
	  \includegraphics[scale=0.5]{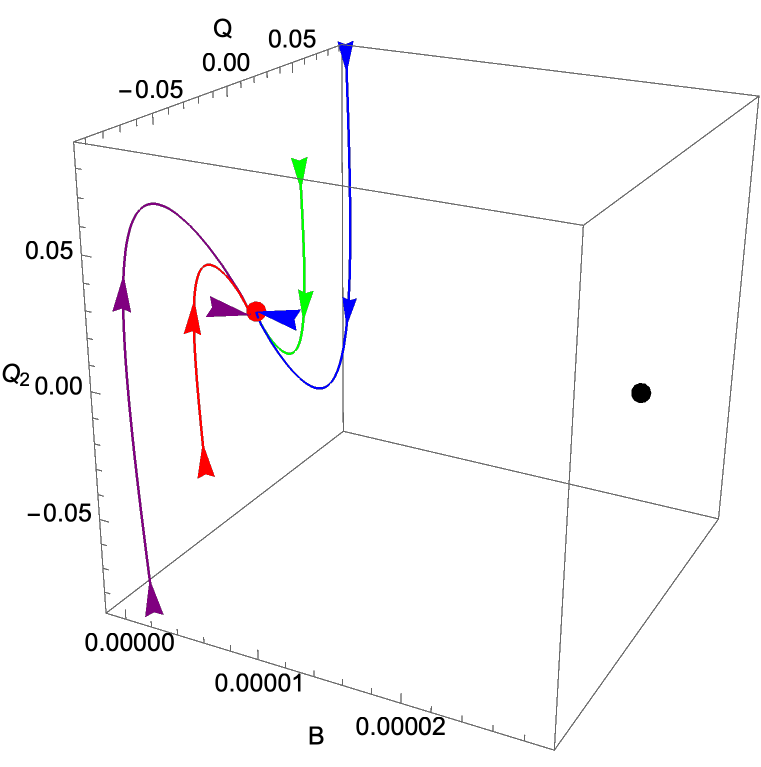}
	
\caption{\it The de Sitter fixed point of the SGECG (displayed as a black point) acts as a repeller rather than an attractor (displayed as a red point).  It is noted that the red point corresponds to a non-de Sitter fixed point. The parameters have been chosen as $\gamma=10^{-5}$, $\alpha=-10^{-10}$, $\kappa=10^{-5}$, $\mu = 10^{-5}$, and $M_p=1$ such that the value of the de Sitter fixed point is $B \simeq 2.82843 \times 10^{-5}$. Different colors correspond to different initial conditions. In particular, the red, green, blue, and purple curves correspond to $\left(B[0],Q[0],Q_2[0] \right)= \left(2 \times 10^{-10}, -4\times 10^{-2}, -4\times 10^{-2} \right)$, $\left(B[0],Q[0],Q_2[0] \right)= \left(2 \times 10^{-10}, 4\times 10^{-2}, 4\times 10^{-2} \right)$, $\left(B[0],Q[0],Q_2[0] \right)= \left(4 \times 10^{-10}, 8\times 10^{-2}, 8\times 10^{-2} \right)$, and $\left(B[0],Q[0],Q_2[0] \right)= \left(4 \times 10^{-10}, -8\times 10^{-2}, -8\times 10^{-2} \right)$, respectively. }
\label{fig2}
\end{figure}

To see if the de Sitter fixed point is an attractor or a repeller to the dynamical system, we numerically solve the dynamical system, following our previous investigations \cite{Do:2023yvg,Pham:2024fub,preprint}. As a result shown in Fig. \ref{fig2}, the de Sitter fixed point is apparently a repeller rather than an attractor, similar to that found in Ref. \cite{preprint}. A non-de Sitter fixed point corresponds to $B=0$, $Q_2=2Q^2$, and the following equation of $Q$,
\begin{equation}
\alpha \left(48+96Q+432Q^2 +1344 Q^3  +384Q^4 \right) + \left(2\kappa+\mu \right) \left( 27Q^2+84Q^3  +24Q^4 \right)  =0.
\end{equation}

It should be noted that our stability analysis is generally valid for the obtained de Sitter solution. In other words, the de Sitter solution of the SGECG is always unstable for $\alpha<0$ as well as $\gamma>0$. In the light of discussions in Refs. \cite{Elizalde:2014xva,Pozdeeva:2019agu,Vernov:2021hxo}, the SGECG seems to be more consistent with the inflationary phase of early universe rather than the accelerated expansion of late-time universe. The reason for this claim is due to the instability of de Sitter inflationary solution, by which the SGECG will not face to the so-called eternal inflation as well as  multiverse scenario \cite{Guth:2007ng}. It has been believed that the so-called graceful exit is the only mechanism to overcome this issue  \cite{Brustein:1994kw}. However, figuring out such a mechanism is not a straightforward task. Before ending this section, it is interesting to recall the fact that the pure Starobinksy model does not admit any exact de Sitter solution but a quasi-de Sitter one \cite{Starobinsky:1980te}. Therefore, admitting no de-Sitter solution or unstable de Sitter solution seems to be a key smoking gun for being a viable inflationary model of any higher-order gravity. 
%%%%%%%%%%%%%%%
 %%%%%%%%%%%%%%%%%%%%%%%%%%
\section{Conclusions}\label{final}
The investigation of whether the GECG with or without the Starobinsky $R^2$ term admits a de Sitter solution as its stable and attractive solution has been carried out. As a result, we have always obtained the same de Sitter solution regardless of the contribution of the $R^2$ term. However, it has been pointed out that the stability of the de Sitter solution cannot be analyzed within the pure GECG due to the lack of information of $\delta \left(\frac{\beta^{(4)}}{\dot\beta^4}\right)$. In other words, the corresponding set of perturbed equations is incomplete such that at least one perturbed field is unconstrained, leaving to the arbitrariness of perturbation modes. It is apparent that the inclusion of the $R^2$ term will resolve easily this issue. In particular, the $R^2$ term will introduce extra terms involving only $\frac{\beta^{(4)}}{\dot\beta^4}$ and will therefore help us to determine explicitly $\delta \left(\frac{\beta^{(4)}}{\dot\beta^4}\right)$ from perturbed field equations, leaving to a complete set of perturbed autonomous equations. As a result, the SGECG with $\alpha<0$ as well as $\gamma>0$ seems to be more suitable for  the inflationary phase of early universe than for the accelerated expansion of late-time universe since it always admits the unstable de Sitter solution. 

In addition to these important findings, we would like to highlight an interesting result for the special case of the GECG, whose field equations become second-order ODEs thanks to a setting  $2\kappa+\mu =-16\alpha$. In particular, this special case admits a stable de Sitter solution rather than an unstable one. Hence, this case seems to be more suitable for the late-time expanding universe than for the early inflationary universe. 

Our stability investigation in the present paper reveals an important cosmological pattern that stable de Sitter solutions tend to emerge from the second-order gravity theories, whereas unstable ones seem to exist in higher-order cases such as fourth-order gravities. In other words, the evolution of our universe likely undergoes a transition from fourth-order field equations at early times to second-order ones at late times. An open question can be asked is that whether third-order field equations play their role in the history of our universe. We leave this interesting question to our future studies. For now, we hope that our current stability analysis could be helpful to related works on fourth-order gravities. We believe that our present paper provides one more example illustrating the important role of the Starobinsky term $R^2$ in higher-order gravities. 
  %%%%%%%%%%%%%%%%
\begin{acknowledgments}
The author would like to thank a referee for his/her constructive comments and suggestions. The author would also like to thank Prof. Phung V. Dong for his supports. This study is funded by the Vietnam National Foundation for Science and Technology Development (NAFOSTED) under grant number 103.01-2023.50.
\end{acknowledgments}
%%%%%%%%%%%%%%%%
\appendix 
\section{Analysis using only the $00$-component of Einstein field equation}\label{app}
In this Appendix, we would like to revisit our investigations presented in sections \ref{sec3} and \ref{sec4} by using another approach used in Ref. \cite{Toporensky:2006kc}, in which only the $00$-component of Einstein field equation is preferred. The reason for this approach is that the $ii$-component of Einstein field equation can be regarded as a differential consequence of the $00$-component, so can be ignored for simplicity. 

Recall the  $00$-component of Einstein field equation, which has been displayed in Eq. \eqref{field-equation-1}
\begin{equation}
\label{App-field-equation-1}
2\dot\beta^2 + \alpha \left(16\dot\beta^6 +144 \dot\beta^2 \ddot\beta^2 -32\ddot\beta^3 +96 \dot\beta \ddot\beta \beta^{(3)} \right) + \left( 2\kappa + \mu\right) \left(9\dot\beta^2 \ddot\beta^2 -2 \ddot\beta^3 +6\dot\beta \ddot\beta \beta^{(3)} \right)=0.
\end{equation}
And shown above, this equation admits the de Sitter solution given by
\begin{equation}
\beta(t) =\zeta t, \quad \zeta =  \left(-\frac{1}{8\alpha} \right)^{\frac{1}{4}}.
\end{equation}
Now, for constructing the corresponding dynamical system we do not need the dynamical variable $Q_2\equiv \frac{\beta^{(3)}}{\dot\beta^3}$ since the field equation is just the third-order ODE. In addition, we still need the other dynamical variables, $B \equiv \frac{1}{\dot\beta^2}$ and $Q\equiv \frac{\ddot\beta}{\dot\beta^2}$ for building up the corresponding autonomous equations, which are given by
\begin{align}
\label{app-dyn-eq-1}
B' & = -2QB,\\
\label{app-dyn-eq-2}
Q' &=\frac{\beta^{(3)}}{\dot\beta^3 } -2Q^2,
\end{align}
where $\frac{\beta^{(3)}}{\dot\beta^3}$ is expected to be figured out  from the field equation \eqref{App-field-equation-1}, which can be rewritten in terms of the introduced dynamical variables $B$ and $Q$ such as
\begin{equation} \label{App-field-equation-2}
2B^2+\alpha \left(16 +144 Q^2 -32 Q^3 +96 Q \frac{\beta^{(3)}}{\dot\beta^3 }   \right) +\left(2\kappa+\mu \right) \left(9Q^2 -2Q^3 +6 Q \frac{\beta^{(3)}}{\dot\beta^3 }  \right)=0.
\end{equation}
It is straightforward to check that $B'=Q'=0$ will admit the de Sitter fixed point with $\frac{\beta^{(3)}}{\dot\beta^3}=Q=0$ and $B^2=-8\alpha$ to this dynamical system. Interestingly, this de Sitter fixed point coincides with that found above in the subsection \ref{fixedpoint}. 

Next, we are going to study the stability of this fixed point. First, we perturb the autonomous equations round the de Sitter fixed point with $\frac{\beta^{(3)}}{\dot\beta^3}=Q=0$ as follows
\begin{align}
\delta B' = -2 \left(Q \delta B + B \delta Q\right) = -2B\delta Q,\\
\delta Q' = \delta \left(\frac{\beta^{(3)}}{\dot\beta^3 }  \right) -4Q \delta Q = \delta \left(\frac{\beta^{(3)}}{\dot\beta^3 }  \right).
\end{align}
The remaining piece we need to define is $ \delta \left(\frac{\beta^{(3)}}{\dot\beta^3 }  \right)$, which should be figured out from the perturbed version of Eq. \eqref{App-field-equation-2}. However, the unexpected situation also happens in this case. It appears in Eq. \eqref{App-field-equation-2} that $\frac{\beta^{(3)}}{\dot\beta^3}$ does not stand alone but couples with $Q$. Consequently, this result will lead to a trouble in defining $ \delta \left(\frac{\beta^{(3)}}{\dot\beta^3 }  \right)$ for the de Sitter fixed point since
\begin{equation}
 \delta \left(Q \frac{\beta^{(3)}}{\dot\beta^3 }  \right) = Q  \delta \left(\frac{\beta^{(3)}}{\dot\beta^3 }  \right)+ \frac{\beta^{(3)}}{\dot\beta^3 } \delta Q =0.
 \end{equation}
 For now, we would like to recall our claim made in the previous section  \ref{sec4} that the stability of the de Sitter fixed point cannot be analyzed due to the lack of information of perturbed variable. 
 
 Very interestingly, the introduction of the Starobinsky term $R^2$ will help us to resolve this issue. The reason is due to the fact that the existence of the $R^2$ term as shown in the action \eqref{SGECG} of the SGECG will introduce an extra term in the corresponding field equation involving $\frac{\beta^{(3)}}{\dot\beta^3}$ without any coupling with $Q$. This feature can be easily seen from Eq. \eqref{field-equation-3}, which can be rewritten in terms of the dynamical variables as follows
 \begin{align} \label{App-field-equation-3}
& 2B^2+ 12\gamma B \left(6 Q -Q^2 +2\frac{\beta^{(3)}}{\dot\beta^3}  \right)+ \alpha \left(16 +144 Q^2 -32 Q^3 +96 Q \frac{\beta^{(3)}}{\dot\beta^3 }   \right) \nonumber\\
& +\left(2\kappa+\mu \right) \left(9Q^2 -2Q^3 +6 Q \frac{\beta^{(3)}}{\dot\beta^3 }  \right)=0.
\end{align}
It is important to note that the existence of the $R^2$ term does not modify the value of the obtained de Sitter solution. However, it does affect on the stability of this de Sitter fixed point, consistent with the discussions in Ref. \cite{Toporensky:2006kc}. As a result, we are able to work out  $ \delta \left(\frac{\beta^{(3)}}{\dot\beta^3 }  \right)$ from the perturbed version of Eq. \eqref{App-field-equation-3},
\begin{align}
\delta \left(\frac{\beta^{(3)}}{\dot\beta^3 } \right)  = -\frac{1}{6\gamma}\left( \delta B +18\gamma \delta Q \right).
\end{align}
Consequently, we now have two perturbation equations of the SGECG  given by
\begin{align}
\delta B' &=  -2B\delta Q,\\
\delta Q' & =  -\frac{1}{6\gamma}\left( \delta B +18\gamma \delta Q \right).
\end{align}
Using exponential perturbations mentioned in Eq. \eqref{expo-pertub} will lead to an equation 
 \begin{equation} \label{App-stability-equation}
 {\cal \bar M}\left( {\begin{array}{*{20}c}
   A_B  \\
   A_Q  \\
 \end{array} } \right) \equiv \left[ {\begin{array}{*{20}c}
   {\lambda} & {2B}    \\
   { \frac{1}{6\gamma}} & {\lambda+3}   \\
    \end{array} } \right]  \left( {\begin{array}{*{20}c}
   A_B  \\
   A_Q  \\
 \end{array} } \right) = 0.
 \end{equation}
 Mathematically, the equation, $\det {\cal \bar M}=0$, will lead to a quadratic equation of $\lambda$,
 \begin{equation}
 \lambda^2 +3\lambda -\frac{B}{3\gamma} =0,
 \end{equation}
 whose roots are solved to be
 \begin{equation}
\lambda_{\pm} = -\frac{1}{2}\left(3 \pm \sqrt{9+\frac{4B}{3\gamma}} \right).
 \end{equation}
Here, we do not have a trivial root $\lambda =0$, in contrast to the stability analysis approach considered in Sect. \ref{sec4}. The appearance  of this trivial root can be interpreted as a consequence of the existence of fourth-order derivatives in the $ii$-component of Einstein field equation. Importantly, the stability of the de Sitter solution remains unaffected with or without the existence of $\lambda=0$.
 
  Thanks to the solution, $B^2=-8\alpha$, these roots can be rewritten as
 \begin{equation} \label{lambda-pm-new}
 \lambda_{\pm} = -\frac{1}{6}\left(9 \pm \sqrt{81-\frac{96\alpha}{\gamma B}} \right),
 \end{equation}
 which is nothing but that shown in Eq. \eqref{lambda-pm} in Sect. \ref{sec4}. This indeed coincides with  our expectation. As discussed in Sect. \ref{sec4}, the de Sitter solution is apparently unstable for $\gamma>0$ and $\alpha<0$ due to the existence of a positive root $\lambda_- >0$. 
 
\begin{figure}[hbtp] 
 \centering
	  \includegraphics[scale=0.6]{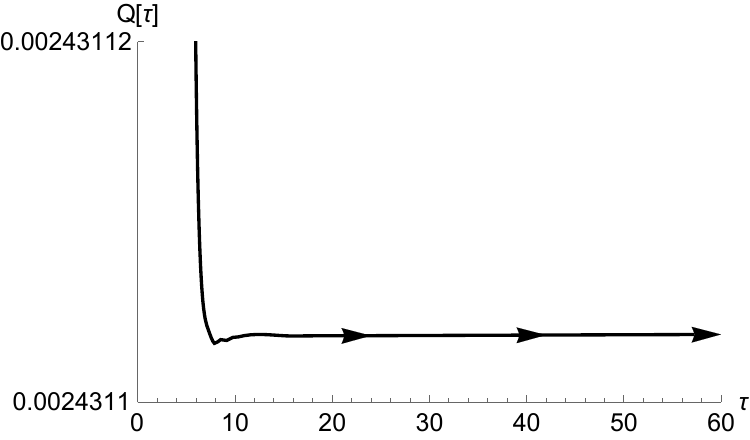}
	
\caption{\it The plot displays the attractive property of $Q \neq 0$ of the non-de Sitter fixed point, implying that the de Sitter fixed point of the SGECG  with $Q = 0$ is a repeller. }
\label{fig3}
\end{figure}

 It is noted that the de Sitter fixed point of the SGECG  can be reverified to be a repeller by solving numerically Eqs. \eqref{app-dyn-eq-1} and \eqref{app-dyn-eq-2} with the help of Eq. \eqref{App-field-equation-3}. Indeed, this statement is confirmed by Fig. \ref{fig3}.
 
In conclusion, it has been shown that the analysis made in this Appendix is really consistent with that done in Sect. \ref{sec4}. This result suggests us a useful hint that using only the $00$-component of Einstein field equation can fully give us solid results. And more importantly, this approach is simpler than the approach using both $00$- and $ii$-components of Einstein field equation as done in the main sections of this paper.
\section{Crosscheck}\label{app2}
In this appendix, we would like to investigate the stability of the obtained de Sitter solution of the GECG using an alternative stability method, which has been used in Ref. \cite{CamposDelgado:2024jst}, where only the $00$-component of the Einstein field equation is considered and the dynamical system is no longer used. Remarkably, this method has been shown to be consistent with the dynamical system method within the framework of the so-called Einstein-Grisaru-Zanon gravity \cite{TQDo}, which is a novel higher-order gravity \cite{CamposDelgado:2024jst}.

Accordingly, we will perturb directly Eq. \eqref{field-equation-1} around the de Sitter solution with $\beta(t) \to \beta(t) +\delta \beta(t)$ and without converting it into autonomous equations of dynamical system. As a result, we obtain the following perturbation equation given by
\begin{align} \label{new-perturbation-equation-app2-0}
\left(1+24 \alpha \zeta^4 \right)  \delta \dot\beta =0.
\end{align}
Furthermore, this equation can be simplified thanks to the found de Sitter solution, i.e., $1+8\alpha \zeta^4=0$, as
\begin{equation}
16 \alpha \zeta^4 \delta \dot\beta =0,
\end{equation}
which gives
\begin{equation}
\delta \dot\beta =0,
\end{equation}
for the non-vanishing $\alpha$ and $\zeta$.
This equation, however, does not tell us the exact behavior of $\delta \beta$. Indeed, if we consider an exponential perturbation such as
\begin{equation}
\delta \beta (t) = C_\beta e^{\hat\lambda \zeta t},
\end{equation}
 then we will have an algebraic equation given by
\begin{equation}
 C_\beta \hat\lambda \zeta e^{\hat\lambda \zeta t}=0,
 \end{equation}
 whose solution is either $C_\beta =0$ or $\hat\lambda =0$ since $\zeta \neq 0$.  Furthermore, if we  perturb Eq. \eqref{field-equation-2}, then we will obtain
 \begin{equation}
 12 \zeta \left(1+24 \alpha \zeta^4 \right)\delta \dot\beta +4\delta \ddot\beta =0,
 \end{equation}
 which can be reduced to
 \begin{equation} 
\delta \ddot\beta =0,
 \end{equation}
 thanks to Eq. \eqref{new-perturbation-equation-app2-0}. 
Interestingly, this equation also admits either $C_\beta =0$ or $\hat\lambda =0$.  According to  these results, we arrive at an important consequence that it is impossible to conclude the stability of the de Sitter solution, consistent with our conclusion obtained in Sect. \ref{sec3} via the dynamical system approach. 

Now, if we impose the role of the Starobinsky term $R^2$ on the field equations of the GECG, we will have, according to Eq. \eqref{field-equation-3}, that
\begin{equation} \label{new-perturbation-equation-app2}
4 \delta \dot\beta +72 \gamma \dot\beta \delta \ddot\beta + 24 \gamma \delta \beta^{(3)} + 96\alpha \dot\beta^4 \delta \dot\beta =0.
\end{equation}
It is transparent that two $\gamma$-terms in this equation will totally change the behavior of the perturbations of $\beta$. Indeed, by still considering the exponential perturbation, i.e., $\delta \beta(t) = C_\beta e^{\hat\lambda \zeta t}$, we will obtain the following algebraic equation from Eq. \eqref{new-perturbation-equation-app2},
\begin{equation} \label{eq-of-hat-lambda}
\hat\lambda \left( 3 \gamma \zeta^2 \hat \lambda^2 +9 \gamma  \zeta ^2 \hat\lambda -1  \right) =0,
\end{equation}  
where the de Sitter solution $1+8\alpha \zeta^4=0$ has been called. 
Obviously, it is straightforward to see that this equation will always admit at least one positive root, i.e., $\hat\lambda >0$, since $\gamma >0$ as required in the pure Starobinsky model. To be more precise, all solutions of Eq. \eqref{eq-of-hat-lambda} are solved to be
\begin{equation} \label{hat-lambda}
\hat\lambda_1 =0, \quad \hat\lambda_{2,3} = \hat\lambda_\pm = - \frac{1}{6} \left[9\pm \frac{\sqrt{3\gamma \left(27 \gamma  \zeta ^2+4\right)}}{\gamma  \zeta}\right].
\end{equation}
It becomes clear that $\hat\lambda_3= \hat\lambda_-$ shown in Eq. \eqref{hat-lambda} is clearly positive definite since $\gamma >0$ as well as $\zeta>0$. Hence, we can conclude that the de Sitter solution of the SGECG is apparently unstable against perturbations. 

More interestingly, $\hat\lambda_{2,3}$ can be rewritten, after some simple algebra,  to be
\begin{equation}
\hat\lambda_{2,3} =\hat\lambda_\pm = - \frac{1}{6} \left[9\pm \sqrt{81 - \frac{96\alpha }{\gamma B }}\right],
\end{equation}
which coincide with $\lambda_\pm$ obtained above, e.g., see Eq. \eqref{lambda-pm-new}. This result does confirm that all calculations carried out through the dynamical system method are valid. 
%%%%%%%%%%%%%%%%%%%%%%%%%%%%%%%%%%%%%%%%%%

%%%%%%%
\end{document}